# A graph theoretic approach for modelling wildlife corridors

Saurabh Shanu[1], Jobin Idiculla[2], Qamar Qureshi[3], Yadvendradev Jhala[3], Sudeepto Bhattacharya[4]

[1]University of Petroleum and Energy Studies, Dehradun, Uttarakhand, India

[2]Central University of Tamil Nadu, Thiruvarur, Tamil Nadu, India

[3]Wildlife Institute of India, Dehradun, Uttarakhand, India

[4]Shiv Nadar University, Gautam Buddha Nagar, Uttar Pradesh, India

**Author for correspondence:**

Sudeepto Bhattacharya, PhD

Department of Mathematics, School of Natural Sciences, Shiv Nadar University,

P.O. Shiv Nadar University, Greater Noida, Gautam Buddha Nagar 201 314, Uttar Pradesh, India

Email: sudeepto.bhattacharya@snu.edu.in

Comments: 32 pages including title page, 10 figures, 1 table.

**Abstract**

Wildlife corridors are components of landscapes, which facilitate the movement of organisms and processes between areas of intact habitat, and thus provide landscape corridor. Corridors are thus regions within a given landscape that generally comprise native vegetation, and connect otherwise fragmented, disconnected, non-contiguous wildlife habitat patches in the landscape. The purpose of designing corridors as a conservation strategy is primarily to counter, and to the extent possible, mitigate the impacts of habitat fragmentation and loss on the biodiversity of the landscape, as well as support continuance of land use for essential local and global economic activities in the region of reference.

In this paper, we use game theory and graph theory to model and design a wildlife corridor network in the Central India – Eastern Ghats landscape complex, with tiger (*Panthera tigris tigris*) as the focal species. We construct a graph using the habitat patches supporting wild tiger populations in the landscape complex as vertices and the possible paths between these vertices as edges. A cost matrix is constructed to indicate the cost incurred by the tiger for passage between the habitat patches in the landscape, by modelling a two-person Prisoner's Dilemma game. A minimum spanning tree is then obtained by employing Kruskal's algorithm, which would suggest a feasible tiger corridor network for the tiger population within the landscape complex. Additionally, analysis of the graph is done using various centrality measures, in order to identify and focus on potentially important habitat patches, and their potential community structure. Correlation analysis is performed on the centrality indices to draw out interesting trends in the network.

**Keywords**: Landscape complex, Corridor, Prisoner's Dilemma game, Graph theory, Minimum spanning tree, Centrality measures, Community detection

**1. Introduction**

Landscape linkage may be defined as the degree to which the landscape impedes or facilitates movement among resource patches [57, 58]. We also define corridor as a habitat, usually linear, embedded in a dissimilar matrix within a landscape, that connects two or more bigger patches of habitat, thereby providing linkage between the habitats and that is proposed for conservation on the grounds that it will enhance or maintain the viability of specific wildlife populations in the habitat patches. Further, we define passage as travel via a corridor by individual animals from one habitat patch to another [7].

Wildlife corridors, as implied from the definition above, are integral components of ecological landscapes. The objective of wildlife corridors is to facilitate the movement of organisms and processes between areas of intact habitat present in the landscape. Corridors are thus regions within a given landscape that generally comprise native vegetation, and connect otherwise fragmented, disconnected, non-contiguous wildlife habitat patches – islands – in the landscape [7, 16, 17].

Landscapes are dynamic and characteristically possess structural (pattern) and functional (process) attributes. Corridors, being integral components of landscapes, are characterized by two distinct categories of components, namely, pattern and process components [16]. The structural corridor between two habitat patches is given by the physical existence of the landscape between the patches. The functional corridor is a product of both – species and landscape. Hence, a functional wildlife corridor is both, a species - as well as landscape-specific concept. Corridors thus, may be considered as emergent phenomena, caused by the interaction between pattern and process attributes of the landscape. The essential function and utility of wildlife corridors is thus to connect at least two distinct habitat areas of biological significance, and thus ensure gene flow between spatially separate populations of species, fragmented due to landscape modifications, by supporting the movements of both biotic and abiotic processes [3, 6, 7, 16, 20, 33, 39, 48, 54, 56].

Scholarship, particularly since the last two decade of the twentieth century continuing till the present, have generally argued in favour of the role of wildlife corridors between fragmented habitat patches. Researchers have demonstrated that presence of species-specific wildlife corridors within a given landscape tobe instrumental in increasing gene flow and population sizes of the species [17, 30, 29, 31, 32].

The above discussions imply that any feasible, realistic modelling to design wildlife corridor must be a species – specific exercise, with a proper choice of habitat for that focal species. In the present paper, we present a computational procedure for designing corridor for the Indian tiger *Panthera tigris tigris* in the Central India-Eastern Ghats landscape complex. For a country biogeographically as vast and diverse as India, relative spatial location of tiger reserves with reference to one another becomes an important attribute to consider for making optimal decision for resource allocations, and thus either protecting existing tiger corridors, or even in some instances, creating proper wildlife corridors in. A key objective in such a decision making therefore would be to select the critical tiger habitats (CTH) in a manner that their spatial configuration ensures a high degree of interconnectivity within often intensely human-dominated landscapes, over a long term land use scenario.

One means to achieve the above objective would be to design the interconnectivity among the existing (or even potential) habitats or CTH using a network model. In such a network, each tiger habitat would be treated as a vertex, and the tiger corridors between these vertices would be the edges.

The present paper comprises a structural study of the aforementioned potential network, through the application of graph theory. The primary objective of this paper is to provide a basic computational framework for perceiving a viable corridor network design in the focal landscape complex for tigers. In this work, connectivity for a given pair of vertices is defined as the number of disjoint paths between them, and hence, all arguments that follow are based on the structural definition of connectivity.

We describe the problem of tiger corridor identification, planning and designing within the landscape as a connection subgraph problem [17]. We next incorporate the conflict of interest between the traversing tiger and the landscape features resultant of primarily anthropogenic modifications, through a two-player non-cooperative, general-sum game. Finally, informed about the possible costs, we provide a spanning tree of the minimum cost, which we claim, could serve as a model corridor.

As a secondary objective of the paper, we aim to identify the most crucial habitat patches constituting tiger populations, and explore their community structure, in an attempt to focus efforts towards efficiently informing their management and conservation programmes. The underlying principle behind this is the fact that the definition of connectivity ultimately relies on the integrity of the vertices, which in this case, happen to be habitat patches. As a result, ensuring the conservation of critical habitat patches is just as important as designing ecological corridors connecting these patches.

The problem of identifying potentially important habitat patches and their underlying community structure is addressed using the approach of graph-theoretic centrality measures [4, 5, 10]. Since the primary problem includes the design of an ecological corridor to ensure the viability of travel by individual animals to and from habitat patches, hence some standard centrality measures, such as degree, betweenness, closeness, eigenvector, and subgraph centralities [8, 12, 21, 25] are computed for this purpose. In addition to this, two intermediate-scale (mesoscale) walk-based centralities [23] based on the rescaling of the subgraph centrality are also calculated to find central vertices which were not indicated by the other classical centralities mentioned.

Although the present work makes reference to a landscape map of the focal complex, it is essentially schematic and semi-emperical in nature. Accordingly, the discussions that follow do not refer to any real-world data as would have been obtained through a GIS routine. Since the work focuses on the presence or absence of corridors linking various tiger habitats in the complex, the distances involved, and the ease of movement for the tiger through these corridors, we are, however, of the opinion that the work could serve as a schema for an informed decision-making by conservationists and wildlife managers in designing real-world corridors.

Section 2 contains a brief description of the Central India – Eastern Ghats landscape complex, followed by section 3 containing the essentials of the mathematical concepts that have been used in this paper. Sections 4 and 5 describe the modelling and the conclusion of the work, respectively.

**2. Central India - Eastern Ghats landscape complex**

A landscape complex is a geographical unit comprising contiguous ecological landscapes (or at least connected in the recent past), that have a potential for exchanging genetic material between the tiger populations inhabiting the forests comprising the complex [36, 37].

Conservation potential, viability and designing have been current in the scholarship since the later part of the last century, in an effort to secure the population of wild tigers in the habitats, still intact in face of large-scale escalation in their habitat fragmentation and loss [28, 40, 41, 42, 43, 52, 53].

To make tiger and related species conservation effective, the researchers from Wildlife Institute of India and the National Tiger Conservation Authority have divided entire India into six landscape complexes based on current tiger occupancy within the critical tiger habitats, existing landscape connections that may serve as tiger corridors, and potential for designing such corridors in the landscape complexes [36, 37, 43].

Central Indian plateau and Eastern Ghats depicted in Fig 1 comprise the focal landscape complex for our present study. This complex supports one of the world's major and significantly healthy tiger populations in the wild, and constitutes two very important tiger habitat landscapes, identified for conservation of wild tiger population in the world [36, 37]. The political units that constitute the complex are the states of Uttar Pradesh, Madhya Pradesh, Bihar, Jharkhand, Chhattisgarh, Odhisha, Andhra Pradesh, Maharashtra, and Rajasthan. Persistent anthropogenic impacts leading to relatively high pressure on the ecosystems owing to economic and allied developmental activities in this natural resource-rich region, even since pre-independence, colonial days continued into the present, and over a period spanning nearly two centuries, has resulted into continual degradation of forests in the landscape complex.

Despite the above factor which may be deemed detrimental to the health of regional biodiversity, the landscape complex, together with three Biosphere Reserves, is the largest tiger occupied area in India, and is home to the largest number of tigers in the country. Also in this complex, various Tiger Conservation Units belonging to levels I, II and III have been identified for according priority status for conservation [28, 36, 37, 43]. Thus, in this landscape complex of significantly high conservation value, the task of maintaining the present tiger habitats and recolonizing the ones that had reported tiger occupancy in the recent past is primarily dependent on the existence of viable tiger corridors available for individual animals to use for dispersal and travelling within the complex. Throughout the paper, we shall treat the immediate past and present tiger occupancy sites equivocally as tiger habitat patches in the landscape.

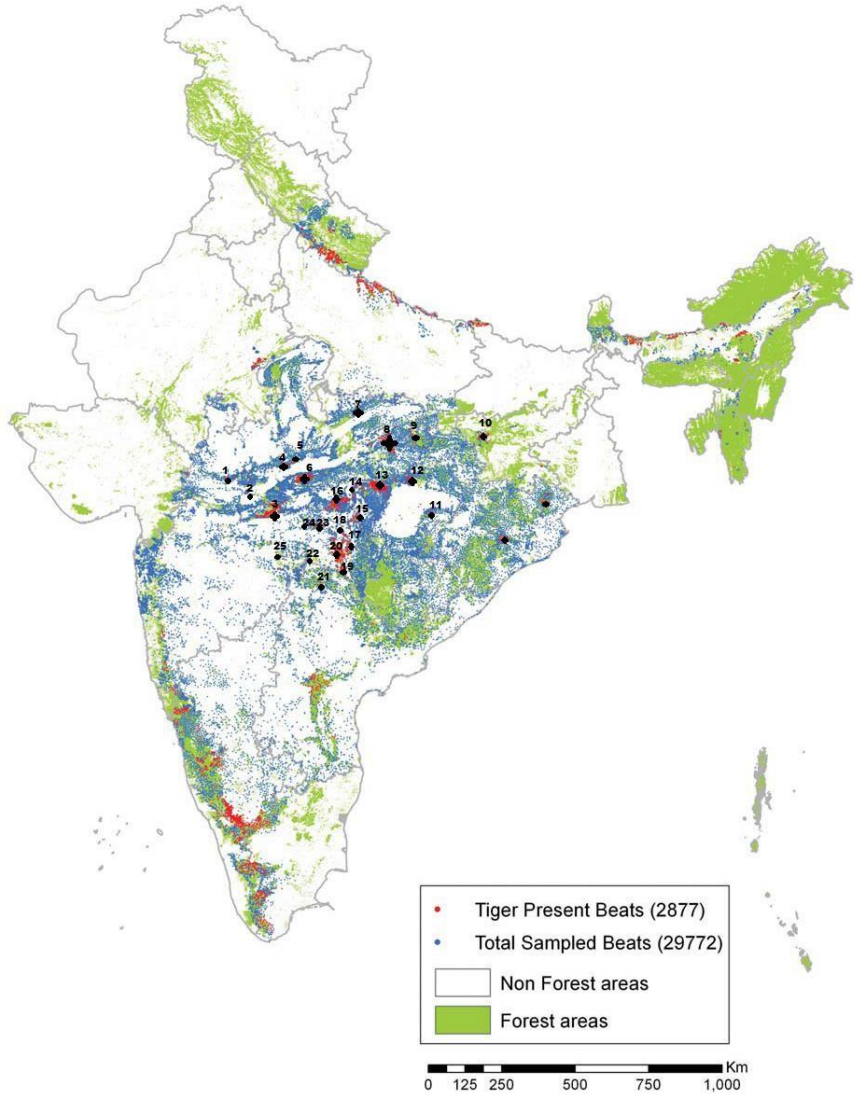

**Fig. 1 Central India-Eastern Ghats landscape complex**

## 3. Game theory, Graph theory, Minimum spanning tree and Centrality measures

In the present work, we shall describe a modelling of a feasible wildlife corridor for the tiger using two specific areas of discrete mathematics. In this section, we shall provide the essentials of both these areas, in order to make the work self-contained.

The first of the two areas mentioned in preceding paragraph is Game theory, which studies and models situations of competition and conflict – of cooperation and defection – between several interacting agents, for shared resources [61]. We use game theory in this paper to model the interactions between possible tiger passages within the landscape, and the different landscape features.

Let $G(\Theta, \Sigma, \Pi)$ be a normal form, strategic game where $\forall i \in I = \{1,...,n\} \subset \mathbb{N}, n \geq 2$,

(i)   $\Theta = \{\Theta_i\}$ is the set of interacting agents or players;

(ii)  $\Sigma_i \neq \{\}$ is the set of strategies for the player $\Theta_i$. $\Sigma = \Sigma_1 \times ... \times \Sigma_n$ is the space of strategies, with $\sigma = (\sigma_1,...,\sigma_n) \in \Sigma$ being a strategy profile of the game $G$;

(iii) $\Pi_i : \Sigma \to \Re$ is the payoff function, which assigns to each strategy profile $\sigma$ a real number $\Pi_i(\sigma)$, the payoff earned by the player $\Theta_i$ when $\sigma$ is played in $G$. $\Pi = \Pi_1 \times ... \times \Pi_n$ is the space of payoff functions in the game.

Let the game $G$ be repeated in periods of discrete time $t \in \aleph$. Assume that the players are 'hardwired' to play only pure strategies in $G$. Thus each strategy set $\Sigma_i$ is a member of the standard basis for the strategy space $\Sigma$ where the $i^{th}$ coordinate is 1 and the rest are zeroes, and thus would correspond to a corner point of the simplex $\Lambda = \left\{ \hat{p} = (p_1, p_2, ..., p_n)^T \in \Re : p_i \geq 0, i \in N, \sum_{i=1}^{n} p_i = 1 \right\}$, which is the simplex corresponding to $\Sigma$ [35].

Let the Prisoner's Dilemma game, a non-zero sum, non-cooperative, symmetric game be represented by $G$ such that $n = 2$. Let the two pure strategies that the two players can opt for, be called cooperate (C) and defect (D), respectively, giving $\Sigma_i = \{C, D\}, i = 1, 2$. The bounded simplex corresponding to $G$ would be given by

$$\Lambda = \left\{ \hat{p} = (p_1, p_2)^T \in \Re : p_i \geq 0, i \in \{1,2\}, \sum_{i=1}^{2} p_i = 1 \right\} \subseteq \Re^2.$$

In the strategic form, $G$ may be described by the following payoff matrix:

|   | C | D |
|---|---|---|
| C | (R,R) | (S,T) |
| D | (T,S) | (P,P) |

With the row player being the first player $\Theta_1$ and the column player being the second one $\Theta_2$
In the above game, both the players $\Theta_1$ and $\Theta_2$ have two pure strategies each to choose from: either play *C* or play *D*. If both play *C*, each obtains a reward *R* as the payoff for cooperating. If both play *D* instead, each obtains a punishment *P* for defecting, as the payoff. If one player plays *C* while the other plays *D*, then the one playing *D* obtains a payoff of temptation (to defect) *T* while the one playing *C* gets a payoff of sucker's, *S*. The game $G$ is then defined by the constraint on the payoffs thus: $T > R > P > S$.

It is obvious from the foregoing discussion, that in a single shot, non-iterated game, the dominant strategy is *D*, and hence both the players, being rational, would choose to play *D* in order to maximize their individual payoffs. However, as the above game matrix shows, in an attempt to maximize individual payoffs, the players obtain equilibrium as $(P, P)$, which, being Nash equilibrium, is a suboptimal solution of the game, the optimal solution being $(R,R)$, that could have been obtained through mutual cooperation of the players. Selfish defection gives a higher payoff than cooperation but if both defect, condition is worse than if both cooperate [35, 61].

Prisoner's Dilemma, though being a general-sum game, would adequately capture and model the essential conflicts of interest among the players involved in the present modelling [1, 2].

The theme of this work is landscape-level conservation planning, given spatial information about locations of entities (tiger habitat patches) of interest. The map in Fig 1 suggests that the possible landscape connectivity between the tiger habitat patches in the focal complex could conveniently be represented as a network. A network is a mathematical model of a real-world situation, which is amenable to analysis by using graph theory, the other area of discrete mathematics that we wish to use for the present modelling [9, 16, 57]. Graph theory has been recognized as a potent framework for modelling landscape connectivity in scholarship, at least since the last decade of the previous century [14, 15, 19, 24, 26, 45, 49, 50, 57, 59, 60]. This body of research serves as our motivation to apply graph theoretic reasoning in the present work to advance our arguments in the paragraphs that follow.

A graph $\Gamma(V(\Gamma), E(\Gamma), \psi_\Gamma)$ (henceforth $\Gamma$) is an ordered triple comprising a set $V(\Gamma)$ of vertices, a set $E(\Gamma)$ of edges, such that $V \cap E = \phi$, and an incidence function $\psi_\Gamma : E \to [V]^2$ where $[V]^2$ is the set of unordered pair of (not necessarily distinct) vertices of $\Gamma$, $\ni$ $e \mapsto \psi_\Gamma(e) = \{v_i, v_j\}, v_i, v_j \in V, \forall e \in E$. The vertices $v_i$ and $v_j$ are incident with the edge $e$, and vice versa. In the aforesaid, the edge $e$ joins the vertices $v_i, v_j$, which, in turn, are the end vertices of $e$. Also, $v_i, v_j$ connected via the incidence function $\psi_\Gamma$, are adjacent to each other. $\Gamma$, as defined thus, is an undirected graph. $\Gamma$ is finite if both $V$ and $E$ are finite sets. Then, $|V|$ the order and $|E|$ the size, define the two parameters of $\Gamma$ respectively. The degree of a vertex $v_i \in \Gamma$ is the number of edges for which $v_i$ is an end vertex. A path in $\Gamma$ is a sequence of vertices $v_1, v_2, ..., v_n$ and a sequence of distinct edges $e_1, e_2, ..., e_{n-1}$ such that each successive pair of vertices $v_k, v_{k+1}$ are adjacent and are the end vertices of $e_k$. A path that begins and ends at the same vertex is a cycle. $\Gamma$ is acyclic if it contains no cycle and is connected if there exists a path from any vertex to any other vertex in $\Gamma$.

A tree $T$ is a connected acyclic graph, and a vertex of the tree that has degree exactly one is a leaf of the tree. If there exists a vertex $v_0 \in T$ such that there exists a unique path from $v_0$ to every other vertex in $T$ but no path from $v_0$ to $v_0$, then $v_0$ is the root of the tree $T$. A tree $T$ is a spanning tree of the connected graph $\Gamma$ if it is a spanning subgraph of $\Gamma$ with vertex set $V(\Gamma)$. We omit the proofs of the following propositions and theorems that we mention for the sake of providing the basis for our arguments and deductions in the paper.

***Proposition 1.*** *In a tree, any two vertices are connected by exactly one path.*

***Proposition 2.*** *Every nontrivial tree has at least two leaves.*

***Theorem 1.*** *If $T(V(T), E(T))$ is a tree, then $|E(T)| = |V(T)| - 1$.*

Let $T$ be a tree in the graph $\Gamma$. If $|V(T)| = |V|$, then $T$ is a spanning tree of $\Gamma$.

***Theorem 2.*** *A graph is connected if and only if it has a spanning tree.*

For the purpose of modelling in the present work, we assume that the tiger habitat patches in the landscape complex constitute the vertices and the collection of landscape connections within this complex that connect any two of the habitats constitutes an edge, comprising the focal landscape complex as a graph $\Gamma(V(\Gamma), E(\Gamma), \psi_\Gamma)$. The existence of an edge between any two vertices represents some ecological flux, such as animal movement, between the adjacent vertices. We shall consider $\Gamma$ being undirected and finite graph in this work, and shall employ the graph distance, which measures the number of edges that constitute a path connecting any two vertices.

A network is a tuple $\Gamma(V_t(\Gamma), E_t(\Gamma), \Psi_{t\Gamma}, A_t)$, with the time parameter $t \in N$, and where $A_t$ is a dynamic algorithm that explores the dyadic ties present in the network and evolves the network in time. Mathematically, a network is a dynamical object owing to the presence of the time parameter, as against a graph, which is an algebraic object. However, this difference apart, the formal, mathematical behaviour of a network is essentially the same as of a graph. A graph, technically, is then a formal representation of a network, and for our purpose of work reported here, both these objects would be treated synonymously.

Centrality measures were originally developed as a fundamental concept in social network analysis [4, 5, 10]. Their scope has tremendously expanded since then and their extensive application to ecological networks has been proven to be quite fruitful [14, 15, 22]. While the intuitive notion of a centrality measure is that it denotes an order of importance on the vertices or edges of a graph by assigning real values to them and ranking them accordingly, there is no commonly accepted mathematical definition of centrality [12]. However, the result of a centrality measure depends on the structure of the graph, as stated in the following definition of a structural index [12]:

**Structural index.** Let $\Gamma_1(V(\Gamma_1), E(\Gamma_1)\Psi_{\Gamma_1})$ and $\Gamma_2(V(\Gamma_2), E(\Gamma_2)\Psi_{\Gamma_2})$ be two undirected, simple graphs and let $\varphi: V(\Gamma_1) \to V(\Gamma_2)$ be an isomorphism between the graphs. Let $X$ represent the set of vertices or edges of $\Gamma_1$, respectively. Then, $s: X \to \mathbb{R}$ is called a structural index if and only if the following condition is satisfied: $\forall x \in X : \Gamma_1 \cong \Gamma_2 \Rightarrow s_{\Gamma_1}(x) = s_{\Gamma_2}(\varphi(x))$, where $s_{\Gamma_1}(x)$ denotes the value of $s(x)$ in $\Gamma_1$, etc.

Minimally, a centrality measure $c: X \to \mathbb{R}$ is required to be a structural index and therefore, induces a partial order on the set X of the graph in consideration. Hence, $x \in X$ is at least as central as $y \in X$ if $c(y) \leq c(x)$.

In this work, we consider $X = V(\Gamma)$, and describe some relevant vertex-level centrality measures. Our discussion includes the 'classical' centrality measures, as well as some additional ones. These measures, thus, produce a vector of centralities for the corridor network $\Gamma$. Specifically, we shall compute the degree centrality ($DC$), eigenvector centrality ($EC$), subgraph centrality ($SC$), positively-scaled subgraph centrality ($SP_i$), negatively-rescaled subgraph centrality ($SN_i$) betweenness centrality ($BC$), and closeness centrality ($CC$), for analysis of the tiger corridor network. Additionally, we shall use the Newman – Girvan algorithm based on edge-betweenness centrality for the purpose of detecting communities in the network. To avoid redundancies in the following discussion, let the graph in consideration be $\Gamma = \Gamma(V(\Gamma), E(\Gamma), \Psi_\Gamma)$, which is undirected. A brief description of each of the aforementioned centralities and the information it yields with regard to the tiger corridor network, follows in the paragraphs below.

**Degree centrality ($DC$).** The degree centrality of a vertex is the number of edges incident to it [25]. In formal notation, degree centrality of a vertex $v_i$,

$$DC(v_i) := \deg(v_i)$$

The significance of the degree centrality is justified by the fact that it assigns a measure of higher importance to vertices which are adjacent to more edges. This makes such vertices representative of patches having alternate pathways to reach other patches, making their usage relatively advantageous, as opposed to vertices adjacent to lesser number of edges [9].

In the present network, high degree of a vertex (habitat patch) indicates a higher number of corridors (even if structural) incident to it, thereby implying a higher opportunity for individual animals to traverse through the vertex. Therefore, the conservation of vertices with a high degree centrality is essential, as any com1promise on such vertices directly affects a large number of individuals using these vertices to traverse the landscape.

**Eigenvector centrality ($EC$).** The eigenvector centrality of a vertex is an extension of the concept of vertex degree centrality [8]. Given a vertex $v_i \in V$ of a graph, all neighbours of it would be given by its neighbourhood $N_{v_i} = \{v_j \in V : \exists e \sim \{v_i, v_j\} \in E\}$. Since all neighbours of $v_i$ usually would not have identical degrees themselves (equally central by DC measure), hence this dispersion in the values of DC in the set $N_{v_i}$ becomes a determinant to a centrality for $v_i$, which is now weighted by the degree centralities of the members of its neighbourhood. This centrality is described as the eigenvector centrality of $v_i$, defined as

$$EC(v_i) := e_1(i),$$

where $e_1(i)$ is the $i^{th}$ component of the eigenvector $e_1$ that corresponds to the largest eigenvalue of the adjacency matrix $A$ [22].

In the tiger corridor network, computing the eigenvector centrality for all vertices helps in identifying those habitat patches which themselves may not have a high number of corridors incident to them, yet are still important by virtue of their immediate neighbouring patches which do. Similar to the degree centrality, damage to such patches adversely affects a considerable number of individuals which rely on using such pathways for their traversal.

**Subgraph centrality (SC).** The subgraph centrality [21] of a vertex $v_i$ is defined as the "sum" of closed walks of different lengths in the networks starting and ending at vertex $v_i$. Formally, if the total number of closed walks of length $k$ originating and terminating at $v_i$ is $\mu_k(v_i)$, then the subgraph centrality of $v_i$ is given by

$$SC(v_i) := \sum_{k=0}^{\infty} \frac{\mu_k(v_i)}{k!} = \sum_{k=0}^{\infty} \frac{(A^k)_{ii}}{k!},$$

where $A^k$ is the $k^{th}$ power of the adjacency matrix of $\Gamma$ [22]. The subgraph centrality is defined in such a way such that the contribution of closed walks decreases as the length of the walks increases. This rule is imposed based on the observation that motifs in real-world networks are small subgraphs [22].

Subgraph centrality is based on the participation of each vertex in all subgraphs in a graph, and has been used successfully in protein-protein interaction networks to determine clear ranking of vertices based on scale-free characteristics [21, 23]. In the tiger corridor network, vertices with a high value of subgraph centrality indicate those habitat patches which belong to a high proportion of closed habitat patch subnetworks. A higher value of subgraph centrality for a certain patch indicates that it would be possible for a species starting out from that patch, to traverse through a large proportion of patches in the landscape, and return to the starting patch using primarily walks of relatively small lengths. From the viewpoint of the species, this makes such patches important as they are highly efficient for round traversal, taking optimal energy expenditure into account.

**Rescaled subgraph centralities ($SP_i, SN_i$).** The rescaled subgraph centralities [23] are a generalization of the subgraph centrality by giving more weights to walks of shorter length (positive rescaling) or more weights to walks of longer length (negative rescaling).

If $A$ is the adjacency matrix of the graph, then the positively rescaled subgraph centrality of a vertex $v_j$ by a factor of $i$, where $i \in \mathbb{N}$, is formally represented by [23],

$$SP_i(v_j) = \sum_{k=0}^{\infty} \frac{(A^k)_{v_j v_j}}{(k+i)!}$$

It can be conveniently rewritten using matrix functions [34] as,

$$SP_i(v_j) = (\psi_i(A))_{v_j v_j}$$

where $\psi_i(A) = \frac{1}{(i-1)!} \int_0^1 e^{(1-i)A} x^{(i-1)} dx$.

On a similar note, the negatively rescaled subgraph centrality of a vertex $v_j$ by a factor of $i$, where $i \in \mathbb{N}$, is formally represented by [23],

$$SN_i(v_j) = (\sum_{s=0}^{i-1} A^s + A^i e^A)_{v_j v_j}$$

The positively and negatively rescaled subgraph centralities are examples of what are termed as "mesoscale" centrality indices [21, 23, 44]. Such indices were primarily defined to extend the principle behind subgraph centrality by allowing one to penalize walk lengths to any arbitrary measure. This allows one to effectively rescale the network in consideration, prioritizing vertices involved in longer closed walks, or conversely, shorter closed walks, as required for analysis.

Rescaling the subgraph centrality allows one to fine-tune it by giving weights to shorter or longer walks as desired, thereby allowing one to systematically assign importance to vertices, by considering their participation at any intermediate scale in the network. With reference to the tiger corridor network, this allows one to identify patches from which an individual animal would be able to move to a large number of other patches and return, by shorter closed walks (shorter than that which permits detection by the subgraph centrality) or by longer closed walks. The latter becomes indispensable in the event that the network in question only has vertices that participate in long closed walks, in which case it makes sense to give more weight to walks of longer length, however, provided that such paths inflict lower mortality risk.

**Betweenness centrality ($BC$).** Betweenness centrality quantifies the number of times a vertex acts as a bridge along the shortest path between two other vertices. Formally, the betweenness centrality of a vertex $v_i$ is given by the total number of shortest paths between all possible pairs of vertices in the graph, that pass through $v_i$, and is defined as

$$BC(v_i) := \sum_{s \neq v_i \neq t} \frac{\sigma_{st}(v_i)}{\sigma_{st}}$$

where $\sigma_{st}$ denotes the number of shortest paths from vertex $s$ to vertex $t$, and $\sigma_{st}(v_i)$ denotes the number of such paths passing through $v_i$.

The betweenness centrality imparts importance to a vertex based on the extent to which it is connected with other vertices which are not connected with each other. Vertices that have a high probability to occur on a randomly chosen shortest path between two randomly chosen vertices have a high value of betweenness centrality. Betweenness centrality shows important vertices that lie on a high proportion of paths between other vertices in the network [12, 18, 51].

With reference to the tiger corridor network, vertices with a high betweenness centrality indicate those habitat patches which most often act as a bridge between other patches that may not be adjacent otherwise. So, any damage to such vertices causes an immediate disruption in the travel of individuals between two habitats, which heavily rely on such bridges as possibly their only means of inter-patch movement.

**Closeness centrality (CC).** Closeness centrality of a vertex $v_i$ is defined as the reciprocal of the sum of geodesic distances (i.e., the shortest path) between $v_i$ and all other vertices. In formal notation,

$$CC(v_i) = \frac{1}{\sum_{t \in V(\Gamma) \setminus \{v_i\}} d_\Gamma(v_i, t)}$$

where $d_\Gamma(v_i, t)$ denotes the geodesic distance between vertex $v_i$ and $t$ (i.e. the number of edges in the shortest path between $v_i$ and $t$).

The closeness centrality is a direct measure of how close a vertex is relative to other vertices, in terms of the shortest path between them, and hence is a measure of the mean distance from a given vertex to other vertices [26, 51].

Closeness centrality indicates important vertices that can communicate quickly with other vertices of the network, as its very definition is based on geodesic (shortest-path) distance. In the tiger corridor network, vertices with high closeness centrality represent habitat patches which are distance-wise nearest to most other habitat patches. Such patches tend to be chosen as stepping stone habitats by most individual tigers, as inhabiting such patches allows them to move to the other habitat patches quite efficiently in terms of travel distance. Also, knowledge of such patches would prove to be advantageous in terms of preventing epidemic spreading among the species members of different patches by quarantining them, and can also serve as a reference for constructing safe human settlements within the network by choosing those vertices with the lowest values of closeness centrality, so as to reduce the frequency of human-animal conflicts.

**Detecting community structure in the network**

In the study of complex networks, a network is said to have community structure if the vertices of the network can be divided into sets so that each set is densely connected internally, with only a few edges between the sets. In the tiger corridor network, communities represent groups of habitat patches which are related by some similar features. Identifying and studying such communities could prove to be instrumental in further deepening our understanding of species' behaviour and their travel patterns [51].

For the purpose of detecting communities in our network, we use the Newman – Girvan algorithm [27], which is based on edge-betweenness centrality.

**Edge-betweenness centrality.** Let $X = E$. The edge-betweenness centrality is the analogue of the standard vertex betweenness centrality, applied to edges. Edge-betweenness centrality quantifies the number of times an edge acts as a bridge along the shortest path between two vertices. Formally, the betweenness centrality of an edge $e_i$ is given by

$$BC(e_i) = \sum_{s \neq t} \frac{\sigma_{st}(e_i)}{\sigma_{st}}$$

where $\sigma_{st}$ denotes the number of shortest paths from vertex $s$ to vertex $t$, and $\sigma_{st}(e_i)$ denotes the number of such paths passing through $e_i$.

The edge-betweenness centrality imparts importance to an edge based on the extent to which it is connected with vertices which are not connected with each other. Edges that have a high probability to occur on a randomly chosen shortest path between two randomly chosen vertices have a high value of edge-betweenness centrality. Edge-betweenness centrality shows important edges that lie on a high proportion of paths between vertices in the network. With reference to the tiger corridor network, edges with a high edge-betweenness centrality indicate those pathways which most often act as a bridge between patches, which may not be adjacent otherwise [51].

**Newman – Girvan algorithm.** The Newman – Girvan algorithm is an algorithm used for detecting communities in networks. It works based on the principle of edge-betweenness centrality. The idea behind the algorithm is that if a network contains communities or groups that are only loosely connected by a few intergroup edges, then all shortest paths between different communities must go along one of those few edges, and such edges will have high edge-betweenness. By removing these edges, the groups are separated from one another to reveal the underlying community structure of the network.

The algorithm is simply stated as follows:

1. Calculate the edge-betweenness for all edges in the graph.
2. Remove the edge with the highest betweenness.
3. Recalculate the edge-betweenness centrality for all edges affected by the removal.
4. Repeat from step 2 until no edges remain.

## 4. Modelling

In consonance with the objective of estimating the presence of a CTH network across the focal landscape complex, the modelling considers only the topology of the network between the different CTHs. Hence, we do not factor information about the spatial aspects and habitat quality of these CTHs into the model, and consider a graph with unweighted vertices to represent the CTH network.

We thus assume that the flux between any two connected habitat patches would be symmetric on the network.

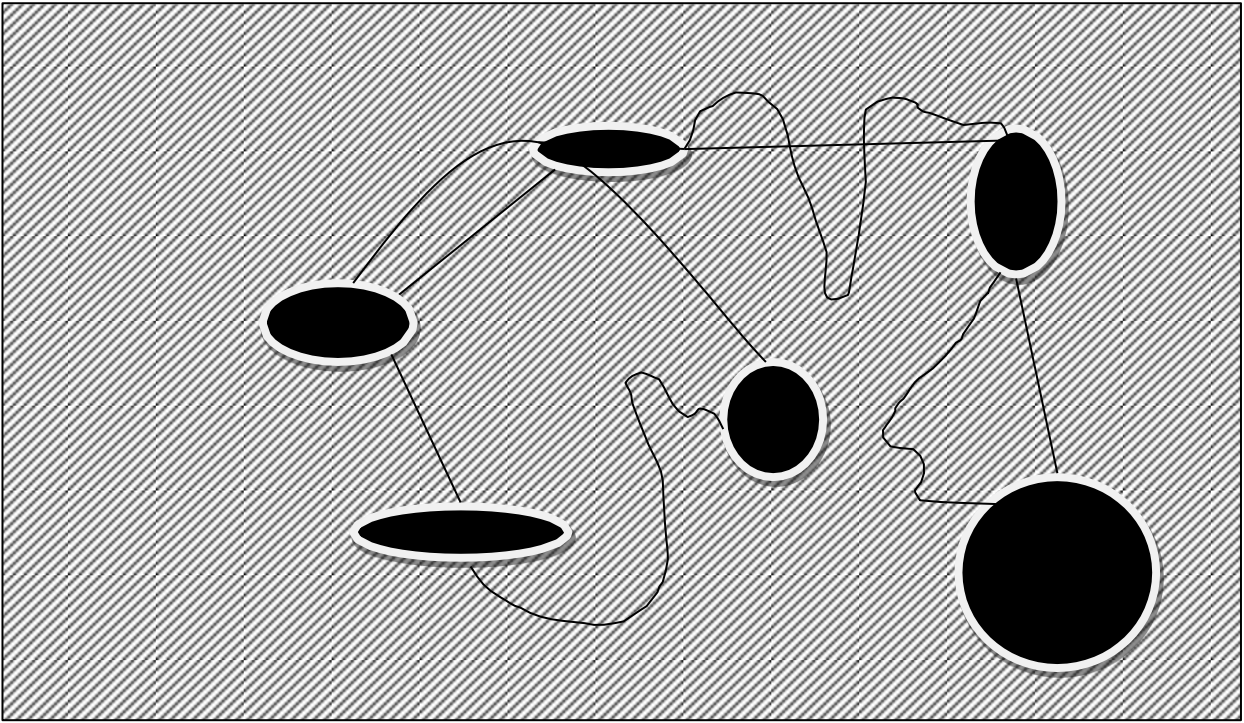

**Fig. 2 Hypothetical landscape showing tiger habitat patches (dark elliptic shapes), corridor between the patches (curves joining the shapes) and the matrix (hashed pattern)**

In Fig. 2, the landscape is represented by a rectangular frame, while the darkened vertices represent habitats for tiger; with the connections between the habitats represented by the lines. The hashed pattern in the figure represents the matrix, a component of the landscape that is neither patch nor corridor in the landscape [16]. The objective for the tiger is to compute a path joining the different habitat patches, which would minimize the risk of its passage through the intervening landscape matrix.

The model is based on the following map depicted in Fig. 4 of the Central India - Eastern Ghats landscape complex, spread over its constituent states [36, 37]:

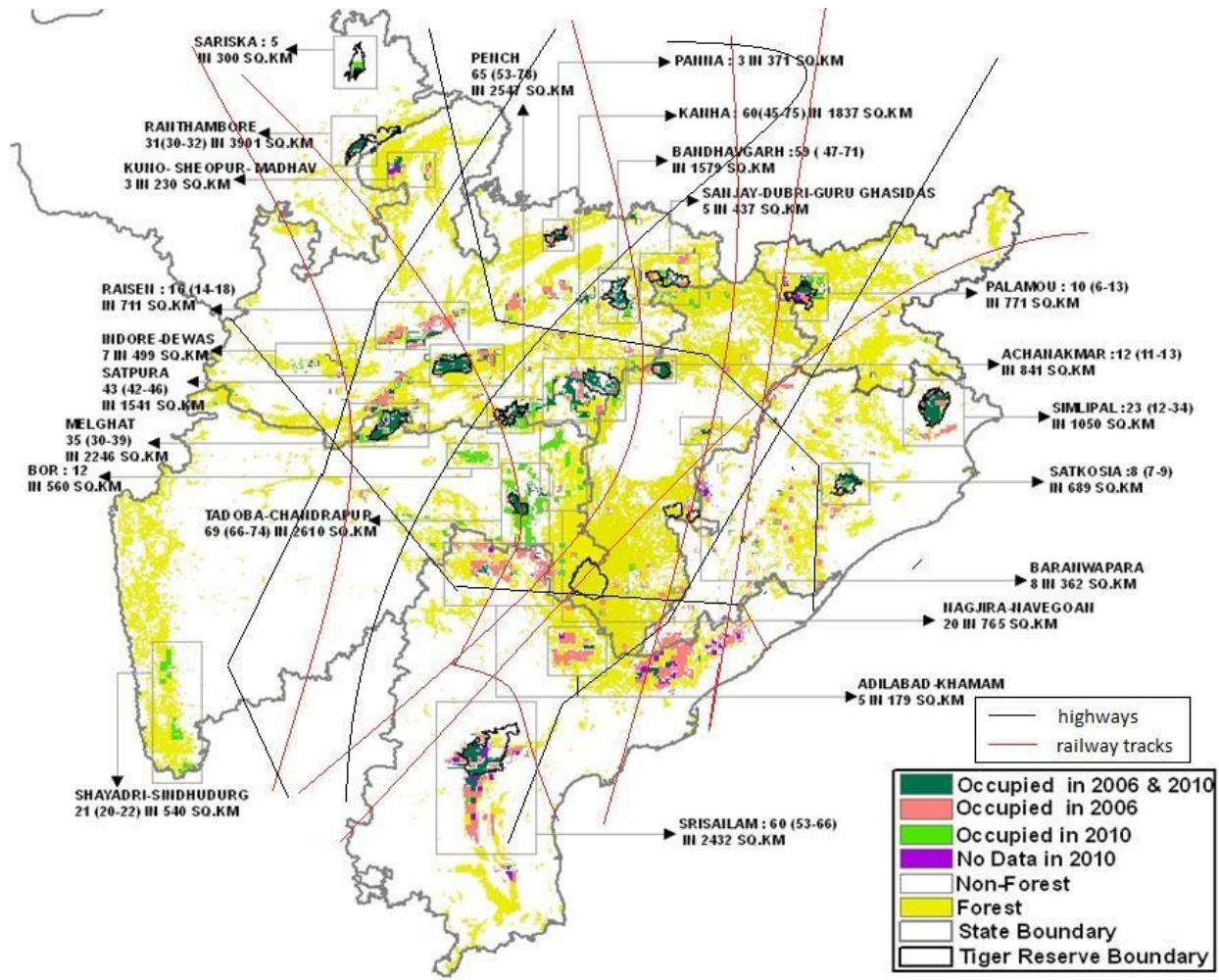

**Fig. 3 Map showing extent and location of tiger habitats in the Central India-Eastern Ghats landscape complex**

To model the possible paths to serve as passage for tigers from a source habitat patch to a destination habitat patch within the landscape complex, we first identify a set of four landscape factors, which may be anthropogenic or natural, and each of which may either promote or constrain the passage of the tiger through the landscape matrix to various degrees, and hence become the major determinants in the structural connections becoming a corridor. The landscape features that we take into consideration are presence of railway tracks (being industrially important, a large number of rail tracks crisscross this region), presence of highways that run through the complex (a good number of major national highways serve the area), presence of forest covers and the absence of forest covers, across the present structural connections between any two tiger populations as given in the map of Fig. 3.

We assume that tiger population in the landscape $(\Theta_1)$ and the set of above mentioned anthropogenic actions and the natural features of the landscape $(\Theta_2)$ constitute the two rational agents that play a two-person Prisoner's Dilemma game $G$ iterated over a number of generations. The players may use a number of strategies in the game in order to optimize their payoff. These payoffs are the costs incurred by the tiger population (called tiger henceforth in the paper) in using the landscape matrix for movement between habitats.

Next we code the different tiger habitats included in the focal landscape complex, as shown in Fig. 3, by the following table:

| S.No | Tiger habitat | Code |
|---|---|---|
| 1. | Sariska | 1 |
| 2. | Ranthambore | 2 |
| 3. | Kuno-Shivpur- Madhav | 3 |
| 4. | Raisen | 4 |
| 5. | Indore-Dewas | 5 |
| 6. | Satpura | 6 |
| 7. | Melghat | 7 |
| 8. | Bor | 8 |
| 9. | Tadoba | 9 |
| 10 | Shayadri | 10 |
| 11 | Srisailam | 11 |
| 12 | Adilabad | 12 |
| 13 | Nagzira | 13 |
| 14 | Baranwapara | 14 |
| 15 | Satkosia | 15 |
| 16 | Simlipal | 16 |
| 17 | Achanakmar | 17 |
| 18 | Palamou | 18 |
| 19 | Sanjay-Dubri-Guru Ghasidas | 19 |
| 20 | Bhandavgrah | 20 |
| 21 | Kanha | 21 |
| 22 | Panna | 22 |
| 23 | Pench | 23 |

Table 1 Coding for the tiger habitats in the complex

Table 1 and the map in Fig. 3 lead to an adjacency matrix $A = [a_{ij}], i = 1,2,...,23; j = 1,2,...,23$ for tiger habitat patches. The matrix is given in Fig. 5 below. The rows, labeled by the index $i$ denote the source habitat $Hi$, while the columns labeled by the index $j$ denote the destination habitats $Hj$ for the tiger. In this paper, since the graph $\Gamma$ is dense and since it is desired to obtain a look-up list to check if there exists a connection (edge) between two tiger habitat patches (vertices), it would be of advantage to compute an adjacency matrix as above. The matrix $A$ stores the presently existing corridor (available contagious forest stretches) between any pair of the twenty three different tiger habitats across the focal landscape, as informed by the map, defined by $a_{ij} = 1$ if the two vertices are connected and $a_{ij} = 0$ otherwise. It may be noted that the storing is effected in $A$ with a time complexity $O(|V|^2)$. Thus, the adjacency matrix *A* describes a landscape corridor network within our focal Central India - Eastern Ghats complex, which constitutes a look-up list to know which all habitats are connected with one another through the existing landscape corridors, thereby describing a planar connected graph for the focal complex.

Matrix *A* in Fig 4 below, describes a landscape connectivity network within the Central India - Eastern Ghats landscape complex through the existing landscape linkages, thereby describing a connected graph for the complex.

|    | 1 | 2 | 3 | 4 | 5 | 6 | 7 | 8 | 9 | 10 | 11 | 12 | 13 | 14 | 15 | 16 | 17 | 18 | 19 | 20 | 21 | 22 | 23 |
|----|---|---|---|---|---|---|---|---|---|----|----|----|----|----|----|----|----|----|----|----|----|----|----|
| 1  | 0 | 1 | 0 | 0 | 0 | 0 | 0 | 0 | 0 | 0  | 0  | 0  | 0  | 0  | 0  | 0  | 0  | 0  | 0  | 0  | 0  | 0  | 0  |
| 2  | 1 | 0 | 1 | 0 | 0 | 0 | 0 | 0 | 0 | 0  | 0  | 0  | 0  | 0  | 0  | 0  | 0  | 0  | 0  | 0  | 0  | 0  | 0  |
| 3  | 0 | 1 | 0 | 1 | 0 | 0 | 0 | 0 | 0 | 0  | 0  | 0  | 0  | 0  | 0  | 0  | 0  | 0  | 0  | 0  | 0  | 1  | 0  |
| 4  | 0 | 0 | 1 | 0 | 1 | 1 | 0 | 0 | 0 | 0  | 0  | 0  | 0  | 0  | 0  | 0  | 0  | 0  | 0  | 0  | 0  | 1  | 0  |
| 5  | 0 | 0 | 0 | 1 | 0 | 1 | 1 | 0 | 0 | 0  | 0  | 0  | 0  | 0  | 0  | 0  | 0  | 0  | 0  | 0  | 0  | 0  | 0  |
| 6  | 0 | 0 | 0 | 1 | 1 | 0 | 1 | 0 | 0 | 0  | 0  | 0  | 0  | 0  | 0  | 0  | 0  | 0  | 1  | 1  | 0  | 0  | 1  |
| 7  | 0 | 0 | 0 | 0 | 1 | 1 | 0 | 1 | 0 | 1  | 0  | 0  | 0  | 0  | 0  | 0  | 0  | 0  | 0  | 0  | 0  | 0  | 1  |
| 8  | 0 | 0 | 0 | 0 | 0 | 0 | 1 | 0 | 1 | 1  | 0  | 0  | 0  | 0  | 0  | 0  | 0  | 0  | 0  | 0  | 0  | 0  | 1  |
| 9  | 0 | 0 | 0 | 0 | 0 | 0 | 0 | 1 | 0 | 1  | 0  | 1  | 1  | 0  | 0  | 0  | 0  | 0  | 0  | 0  | 0  | 0  | 1  |
| 10 | 0 | 0 | 0 | 0 | 0 | 0 | 1 | 1 | 1 | 0  | 0  | 1  | 0  | 0  | 0  | 0  | 0  | 0  | 0  | 0  | 0  | 0  | 0  |
| 11 | 0 | 0 | 0 | 0 | 0 | 0 | 0 | 0 | 0 | 0  | 0  | 1  | 0  | 0  | 0  | 0  | 0  | 0  | 0  | 0  | 0  | 0  | 0  |
| 12 | 0 | 0 | 0 | 0 | 0 | 0 | 0 | 0 | 1 | 1  | 1  | 0  | 1  | 0  | 0  | 0  | 0  | 0  | 0  | 0  | 0  | 0  | 0  |
| 13 | 0 | 0 | 0 | 0 | 0 | 0 | 0 | 0 | 1 | 0  | 0  | 1  | 0  | 1  | 0  | 0  | 0  | 0  | 0  | 1  | 0  | 1  |    |
| 14 | 0 | 0 | 0 | 0 | 0 | 0 | 0 | 0 | 0 | 0  | 0  | 0  | 1  | 0  | 1  | 0  | 1  | 1  | 0  | 0  | 1  | 0  | 0  |
| 15 | 0 | 0 | 0 | 0 | 0 | 0 | 0 | 0 | 0 | 0  | 0  | 0  | 0  | 1  | 0  | 1  | 0  | 0  | 0  | 0  | 0  | 0  | 0  |
| 16 | 0 | 0 | 0 | 0 | 0 | 0 | 0 | 0 | 0 | 0  | 0  | 0  | 0  | 0  | 1  | 0  | 0  | 1  | 0  | 0  | 0  | 0  | 0  |
| 17 | 0 | 0 | 0 | 0 | 0 | 0 | 0 | 0 | 0 | 0  | 0  | 0  | 0  | 1  | 0  | 0  | 0  | 1  | 1  | 1  | 1  | 0  | 0  |
| 18 | 0 | 0 | 0 | 0 | 0 | 0 | 0 | 0 | 0 | 0  | 0  | 0  | 0  | 1  | 0  | 1  | 1  | 0  | 1  | 0  | 0  | 0  | 0  |
| 19 | 0 | 0 | 0 | 0 | 0 | 0 | 0 | 0 | 0 | 0  | 0  | 0  | 0  | 0  | 0  | 0  | 1  | 1  | 0  | 1  | 0  | 0  | 0  |
| 20 | 0 | 0 | 0 | 0 | 0 | 1 | 0 | 0 | 0 | 0  | 0  | 0  | 0  | 0  | 0  | 0  | 1  | 0  | 1  | 0  | 0  | 1  | 0  |
| 21 | 0 | 0 | 0 | 0 | 0 | 1 | 0 | 0 | 0 | 0  | 0  | 0  | 1  | 1  | 0  | 0  | 1  | 0  | 0  | 0  | 0  | 1  | 1  |
| 22 | 0 | 0 | 1 | 1 | 0 | 0 | 0 | 0 | 0 | 0  | 0  | 0  | 0  | 0  | 0  | 0  | 0  | 0  | 0  | 0  | 1  | 1  | 0  |
| 23 | 0 | 0 | 0 | 0 | 0 | 1 | 1 | 1 | 1 | 0  | 0  | 0  | 1  | 0  | 0  | 0  | 0  | 0  | 0  | 0  | 1  | 0  | 0  |

**Fig. 4 Adjacency matrix $A = [a_{ij}]$ for tiger habitats in the Central India-Eastern Ghats landscape complex**

The graph Γ may then be represented pictorially from the adjacency matrix *A*, describing the existing connectivity between the various tiger populations as coded in Table 1:

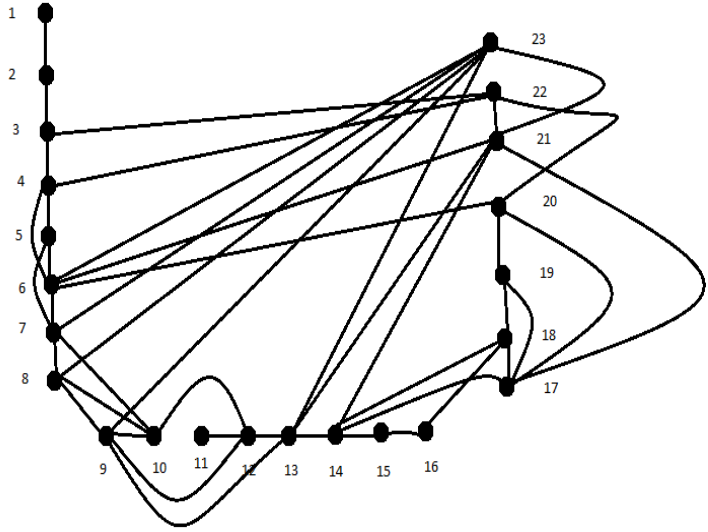

**Fig. 5 The habitat connectivity graph obtained from the adjacency matrix $A$**

We next compute the costs incurred on tigers in using the connections between different populations in the given landscape complex. With each possible edge (corridor) between any two vertices (habitats), we associate a numeric weight *c*, thus rendering Γ a weighted graph. We designate the weight assigned to an edge (corridor) as the cost incurred by the tiger for passage through that corridor, and define this cost function as the mapping

$c : E \to \aleph$
$\ni e \mapsto c(e) = r \in \aleph, \forall e \in E, \aleph = \{0,1,...\}.$

We assume that the cost of using a corridor is a numeric proxy for the perceived (by the tiger) mortality or (even physical) risk associated with the corridor, and hence to the risk incurred by the tiger in traversing that corridor. We further assume that the risks being essentially and only incurred due to the presence or the absence of even one or all, of the above mentioned landscape features.

The costs to each of the possible corridor is assigned taking into consideration the possible kind of features mentioned in the foregoing, that a traversing tiger is likely to encounter while negotiating that corridor. The payoff matrix for the game *G* is constructed based on these costs. One of the prime objectives in designing tiger corridors would be to minimize the risk(mortality or physical injury) or cost objective  to minimize this risk or cost incurred by the tiger in using a particular corridor as a  passage, we describe the research problem as: *Given an undirected, connected graph* $\Gamma(V, E, \psi_\Gamma)$, *an index set* $I = \{0,1,...,n\} \subseteq \aleph$, $\ni v_i \in V, e_i \in E$ *with* $i \in I$, *the cost function* $c_i = c(e_i) \forall i \in I$, *compute a spanning tree T such that* $\sum_{i \in N} c_i$ *is minimum.*

Therefore, the objective of our work is to compute a spanning tree for the given graph, such that the sum total of the costs incurred by the tiger in its passage between the habitat patches embedded in the given landscape complex, through the landscape matrix, is minimized.

One of the most commonly used solution procedure to address the research problem is the Boruvka-Kruskal algorithm (Kruskal's algorithm) [11, 46, 47]. Kruskal's algorithm is a tree-search algorithm that accepts as input a weighted connected graph, and returns as output an optimal spanning tree. The execution of the Algorithm starts with $|V|$ isolated trees in the forest (a set of trees, and hence essentially an acyclic graph), each initially with 1 vertex. The Algorithm then constructs a spanning tree edge-by-edge, by making a decision to select the least cost path that connects two trees, to return a single tree in the forest. At the termination of the Algorithm, the forest has only 1 component, namely, the output spanning tree. Being a greedy algorithm, Kruskal's algorithm makes a 'greedy' (locally optimal) decision at each stage of its run, without being concerned about the impact of this decision on the global optimality of the output.

A major advantage of using Kruskal's algorithm for solving our defined problem is that the Algorithm has a linear time complexity, given by $O(|E|\log|E|)$. Additionally, for Kruskal's Algorithm, the following theorem guarantees the optimality of the output spanning tree:

***Theorem 3*** *Every Bourvka-Kruskal tree is an optimal tree* (Bondy and Murty 2008).

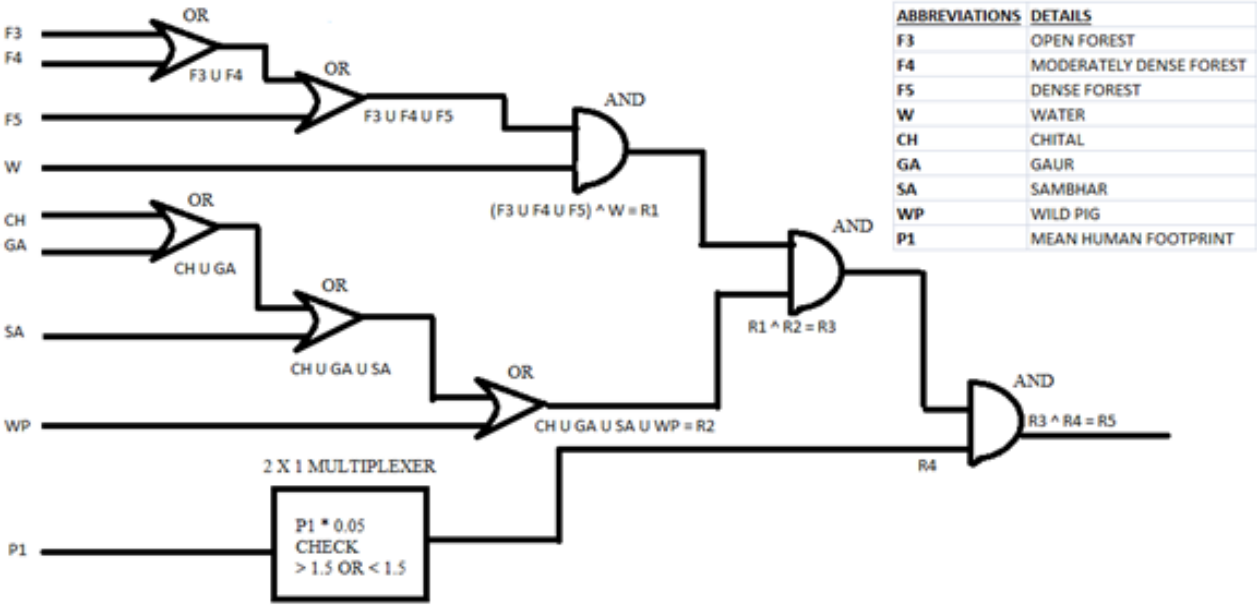

**Fig.6 Logic gate circuit for the decision flow for evaluation of a viable linkage between two habitat patches.**

The cost $c(e_i) = c_i, i \in I$ incurred by the tiger in using edge (corridor) $e_i$ between any two of the twenty-three habitats of Table 1 are evaluated using the payoff matrix of the two-player Prisoner's Dilemma game and a logic gate circuit, as has already been remarked in the above discussion and Fig.6. Using the logic gate circuit, we compute the overall payoff of all the parameters together. The OR gate which computes the union between the components has been applied for the factors which exist in a grid either as a singleton component or a combination of components. The AND gate which computes the intersection between the components has been applied for the factors which need to coexist for facilitating the movement of the tigers through the grid. A 2X1 multiplexer is used as we intend to check the interaction between two commodities (tiger and the interacting factor while modelling the game) and obtain a single value either **0 or 1** based on a threshold value. A combination of all the discussed logic gates is used to obtain a final circuit which evaluates:

(i) The combination of few factors out of which presence of any one type, would facilitate the movement of tigers e.g. forest types.
(ii) The combination of few factors out of which presence of all types, would facilitate the movement of tigers e.g. prey base and water presence.
(iii) The values of presence or absence of any factor based on a particular threshold value e.g. anthropogenic inputs.

In computing the payoff matrix, we further assume that the players involved in this game only choose to play pure strategies. The reasoning for various landscape features that we consider as impacts on tiger corridor in the landscape complex, and their corresponding cost assignments and subsequent payoff evaluations are as below:

1. Highways: tigers can have two distinct approaches while negotiating a highway during a passage from the source to the destination habitat patch. They may either choose to move along or through the highway, thus 'cooperate' with the existence of the highway and play a strategy C, and thereby run a risk of incurring a very heavy cost to itself, often resulting in fatality, or may choose to avoid the highway and try to look for another possible path to the target habitat patch, and thus play the strategy D. However, in order to reach a rich habitat patch situated diametrically on the opposite side of a highway, the tiger would have no strategic alternative but necessarily has to cross the highway and reach the target patch. On the other hand, the highways always defect with the movements and path choice of tigers, and thus play ALL D. So in summary, in such a scenario as above, the tigers act as non-cooperators and the highways as defectors in the game and thus the net risk score due to highways are given as 5 units.
2. Railways: Very similar to the highways, the rail tracks laid between the corridors incur a risk to the tiger by being present in the landscape matrix, and we assign the factor a risk cost of 5 units to the tiger.
3. Forest cover: This feature, if present in adequacy, acts to benefit the movements of tigers by way of providing shelter, and we assume, prey base during transition through the covers. So the forest covers act as a cooperator to the tigers (that is, play C) and thus add to the benefit and reduce the passage risk cost by an amount of 3 units, or, contributes a quantum of -3 units to the cost incurred by the tiger.
4. Absence of forest cover: when the forest cover is not present between the source and the destination habitat patches in the landscape, then passage through such a matrix enhances the risk for the tiger by acting in essence opposite to tiger's interest, thereby incurring a cost of 5 units to the transient tiger.

Based on the above criteria of scoring, the various factors with respect to tiger using the strategy pair of Prisoner's Dilemma game the following cost matrix is obtained, with scores for the edges entered in the matrix only when there does exist an edge in Fig. 5, connecting any two of the vertices of the graph:

| Patch | 1 | 2 | 3 | 4 | 5 | 6 | 7 | 8 | 9 | 10 | 11 | 12 | 13 | 14 | 15 | 16 | 17 | 18 | 19 | 20 | 21 | 22 | 23 |
|---|---|---|---|---|---|---|---|---|---|---|---|---|---|---|---|---|---|---|---|---|---|---|---|
| 1 |   | 5 |   |   |   |   |   |   |   |   |   |   |   |   |   |   |   |   |   |   |   |   |   |
| 2 | 5 |   | 3 |   |   |   |   |   |   |   |   |   |   |   |   |   |   |   |   |   |   |   |   |
| 3 |   | 3 |   | 12 |   |   |   |   |   |   |   |   |   |   |   |   |   |   |   |   |   | 12 |   |
| 4 |   |   | 12 |   | 7 | 10 |   |   |   |   |   |   |   |   |   |   |   |   |   |   |   | 7 |   |
| 5 |   |   |   | 7 |   | 12 | 12 |   |   |   |   |   |   |   |   |   |   |   |   |   |   |   |   |
| 6 |   |   |   | 10 | 12 |   | 3 |   |   |   |   |   |   |   |   |   |   |   |   | 12 | 7 |   | 7 |
| 7 |   |   |   |   | 12 | 3 |   | 7 |   | 20 |   |   |   |   |   |   |   |   |   |   |   |   | 2 |
| 8 |   |   |   |   |   |   | 7 |   | 2 | 20 |   |   |   |   |   |   |   |   |   |   |   |   | 7 |
| 9 |   |   |   |   |   |   |   | 2 |   | 17 |   | 2 | 3 |   |   |   |   |   |   |   |   |   | 10 |
| 10 |   |   |   |   |   | 20 | 20 | 17 |   |   |   | 17 |   |   |   |   |   |   |   |   |   |   |   |
| 11 |   |   |   |   |   |   |   |   |   |   |   | 30 |   |   |   |   |   |   |   |   |   |   |   |
| 12 |   |   |   |   |   |   |   |   | 2 | 17 | 30 |   | 7 |   |   |   |   |   |   |   |   |   |   |
| 13 |   |   |   |   |   |   |   |   | 3 |   |   | 7 |   | 12 |   |   |   |   |   |   | 3 |   | 3 |
| 14 |   |   |   |   |   |   |   |   |   |   |   |   | 12 |   | 27 |   | 25 | 22 |   |   | 15 |   |   |
| 15 |   |   |   |   |   |   |   |   |   |   |   |   |   | 27 |   | 2 |   |   |   |   |   |   |   |
| 16 |   |   |   |   |   |   |   |   |   |   |   |   |   |   | 2 |   | 7 |   |   |   |   |   |   |
| 17 |   |   |   |   |   |   |   |   |   |   |   |   |   | 22 |   |   | 12 | 7 | 7 | 2 |   |   |   |
| 18 |   |   |   |   |   |   |   |   |   |   |   |   |   | 22 |   | 7 | 12 |   | 7 |   |   |   |   |
| 19 |   |   |   |   |   |   |   |   |   |   |   |   |   |   |   |   | 7 | 7 |   | 2 |   |   |   |
| 20 |   |   |   |   | 12 |   |   |   |   |   |   |   |   |   |   |   | 7 |   | 2 |   |   | 7 |   |
| 21 |   |   |   |   | 7 |   |   |   |   |   |   |   | 3 | 15 |   |   | 2 |   |   |   |   | 12 | 3 |
| 22 |   |   | 12 | 7 |   |   |   |   |   |   |   |   |   |   |   |   |   |   |   | 7 | 12 |   |   |
| 23 |   |   |   |   |   | 7 | 2 | 7 | 10 |   |   |   | 3 |   |   |   |   |   |   |   | 3 |   |   |

**Fig. 7 Cost matrix of the tiger for using existing corridors between different habitat patches in the complex**

The pseudo code for Kruskal's algorithm for generating the minimum spanning tree is as below:

*Procedure Kruskal*$(\Gamma, c)$

    START
    DECLARE

E = set of Edges
A = vertex value
F = storage set for edges
a, b = initial vertices with minimum cost
n = number of vertices
e = edge between two considered vertices
  BEGIN
  $F := E; A = \phi$
  Set initial e = min (all the weights)
  Set the vertices containing the initial cost as the initial vertices $(a-b)$
  while $|A| < n-1$ loop
      find $e \in F \ni c(e)$ is minimum
      $F := F - \{e\}$
      if $\Gamma(A \cup \{e\})$ acyclic then
        $A := A \cup \{e\};$
      end if;
  end loop
  $\Gamma(A)$ is a minimum spanning tree end Kruskal;
  END

A minimum spanning tree (MST) for the focal complex, obtained on using Kruskal's algorithm, with its optimality guaranteed by Theorem 3, is shown in Fig 7:

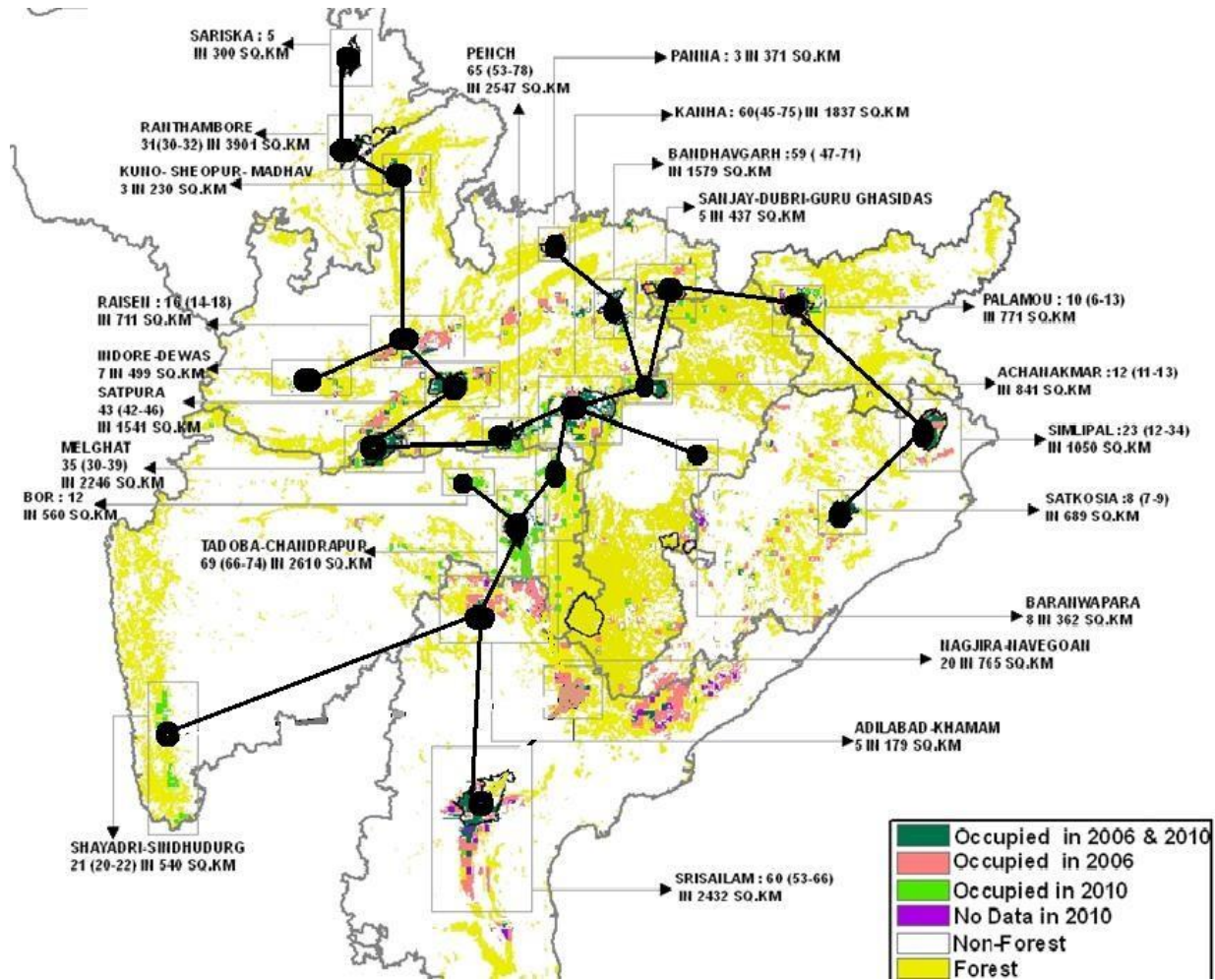

**Fig. 8 A feasible tiger corridor network, given by a MST using Kruskal's algorithm, overlaid on the map of the focal landscape complex**

This tree, by virtue of Theorem 2, guarantees the connectivity of corridors in the focal landscape complex, represented in this work by the graph $\Gamma$, and now by the spanning tree. Further, Theorem 3 ensures that the tree obtained through the algorithm is indeed a MST.

Centrality analysis is performed on the network to identify potentially important patches. The following centralities are computed: Degree centrality ($DC$), Eigenvector centrality ($EC$), Betweenness centrality ($BC$), Closeness centrality ($CC$), Subgraph centrality ($SC$), Positively-rescaled subgraph centrality ($SP_i$) and Negatively-rescaled subgraph centrality ($SN_i$). The computations were performed using algorithms and functions scripted in *MATLAB*. A table of vertices (habitat patches) ranked (highest-to-lowest) by various centrality measures is given below:

| DC | EC | BC | CC | SC | $SP_1$ | $SP_2$ | $SP_3$ | $SP_4$ | $SP_5$ | $SP_6$ | $SN_1$ | $SN_2$ | $SN_3$ | $SN_4$ | $SN_5$ | $SN_6$ |
|---|---|---|---|---|---|---|---|---|---|---|---|---|---|---|---|---|
| 6 | 23 | 21 | 21 | 23 | 23 | 21 | 21 | 21 | 21 | 21 | 23 | 23 | 23 | 23 | 23 | 23 |
| 21 | 21 | 3 | 6 | 21 | 21 | 6 | 6 | 6 | 6 | 6 | 21 | 21 | 21 | 21 | 21 | 21 |
| 23 | 6 | 14 | 13 | 6 | 6 | 20 | 20 | 20 | 20 | 20 | 6 | 6 | 6 | 6 | 6 | 6 |
| 7 | 13 | 13 | 23 | 9 | 9 | 22 | 22 | 22 | 22 | 22 | 9 | 9 | 13 | 13 | 13 | 13 |
| 9 | 9 | 22 | 14 | 13 | 13 | 23 | 17 | 17 | 17 | 17 | 13 | 13 | 9 | 9 | 9 | 9 |
| 13 | 7 | 6 | 22 | 7 | 7 | 14 | 14 | 4 | 4 | 4 | 7 | 7 | 7 | 7 | 7 | 7 |
| 14 | 8 | 2 | 20 | 17 | 17 | 13 | 7 | 7 | 7 | 7 | 8 | 8 | 8 | 8 | 8 | 8 |
| 17 | 14 | 12 | 17 | 8 | 14 | 7 | 13 | 13 | 18 | 18 | 17 | 17 | 17 | 14 | 14 | 14 |
| 4 | 10 | 20 | 7 | 14 | 8 | 17 | 4 | 18 | 13 | 13 | 14 | 14 | 14 | 17 | 17 | 10 |
| 8 | 17 | 4 | 5 | 10 | 10 | 18 | 23 | 14 | 14 | 14 | 10 | 10 | 10 | 10 | 10 | 17 |
| 10 | 12 | 7 | 9 | 12 | 12 | 4 | 18 | 23 | 23 | 10 | 12 | 12 | 12 | 12 | 12 | 12 |
| 12 | 20 | 23 | 4 | 20 | 18 | 9 | 9 | 10 | 10 | 9 | 20 | 20 | 20 | 20 | 20 | 20 |
| 18 | 5 | 17 | 12 | 4 | 4 | 12 | 10 | 9 | 9 | 23 | 4 | 4 | 22 | 22 | 5 | 5 |
| 20 | 22 | 9 | 8 | 18 | 20 | 10 | 12 | 15 | 16 | 19 | 18 | 22 | 4 | 5 | 22 | 22 |
| 22 | 4 | 5 | 19 | 22 | 22 | 8 | 15 | 12 | 15 | 16 | 22 | 18 | 5 | 4 | 4 | 4 |
| 3 | 18 | 15 | 3 | 5 | 5 | 3 | 16 | 16 | 19 | 15 | 5 | 5 | 18 | 18 | 18 | 18 |
| 5 | 19 | 10 | 10 | 19 | 19 | 5 | 19 | 19 | 12 | 12 | 19 | 19 | 19 | 19 | 19 | 19 |
| 19 | 3 | 18 | 18 | 3 | 3 | 19 | 3 | 8 | 8 | 8 | 3 | 3 | 3 | 3 | 3 | 3 |
| 2 | 15 | 8 | 15 | 15 | 15 | 15 | 8 | 3 | 3 | 3 | 15 | 15 | 15 | 15 | 15 | 15 |
| 15 | 11 | 1 | 2 | 16 | 16 | 16 | 5 | 5 | 5 | 5 | 16 | 16 | 16 | 16 | 16 | 16 |
| 16 | 16 | 11 | 11 | 2 | 2 | 2 | 2 | 2 | 2 | 11 | 2 | 11 | 11 | 11 | 11 | 11 |
| 1 | 2 | 16 | 16 | 11 | 11 | 11 | 11 | 11 | 11 | 2 | 11 | 2 | 2 | 2 | 2 | 2 |
| 11 | 1 | 19 | 1 | 1 | 1 | 1 | 1 | 1 | 1 | 1 | 1 | 1 | 1 | 1 | 1 | 1 |

**Fig. 9 Ranking of tiger habitats by various centrality measures**

From the centrality analysis conducted on the network, and by considering the five top-ranking vertices in each centrality, we observe that the most central patches are 3, 6, 7, 9, 13, 14, 17, 20, 21, 22, and 23.

Among these, patches 6, 21, and 23 are highly central and were detected by all the centrality indices except $SP_i$ and $BC$. We also observe that 13 is an important patch as well, as it was undetected only by $SP_i$ and DC. Similarly, 9 was undetected only by $BC, CC,$ and $SP_i$.

The five lowest-ranked vertices according to $CC$ are 1, 16, 11, 2 and 15. These may be considered the most suitable patches for human settlements, so as to minimize human confrontations with the species.

It is interesting to note that certain patches were detected only by select centralities; 7 by $DC$; 3 and 14 by $BC$; 17 and 20 by $SP_i$. This further justifies the use of multiple centrality indices on a network in order to identify the maximum number of important vertices, pertaining to varied perspectives of "importance".

Performing correlation analysis [38] using Pearson coefficients on all the centrality indices, we obtain the following correlation table:

|     | DC    | EC    | BC    | CC    | SC    | $SP_1$ | $SP_2$ | $SP_3$ | $SP_4$ | $SP_5$ | $SP_6$ | $SN_1$ | $SN_2$ | $SN_3$ | $SN_4$ | $SN_5$ | $SN_6$ |
|-----|-------|-------|-------|-------|-------|--------|--------|--------|--------|--------|--------|--------|--------|--------|--------|--------|--------|
| DC  | 1     |       |       |       |       |        |        |        |        |        |        |        |        |        |        |        |        |
| EC  | -0.16 | 1     |       |       |       |        |        |        |        |        |        |        |        |        |        |        |        |
| BC  | -0.28 | -0.1  | 1     |       |       |        |        |        |        |        |        |        |        |        |        |        |        |
| CC  | 0.044 | -0.24 | -0.01 | 1     |       |        |        |        |        |        |        |        |        |        |        |        |        |
| SC  | -0.05 | 0.493 | -0.23 | 0.054 | 1     |        |        |        |        |        |        |        |        |        |        |        |        |
| $SP_1$ | 0.046 | 0.521 | -0.3  | 0.122 | 0.96  | 1      |        |        |        |        |        |        |        |        |        |        |        |
| $SP_2$ | -0.08 | -0.11 | 0.182 | 0.449 | 0.112 | 0.054  | 1      |        |        |        |        |        |        |        |        |        |        |
| $SP_3$ | -0.21 | 0.278 | -0.03 | 0.297 | 0.199 | 0.258  | 0.335  | 1      |        |        |        |        |        |        |        |        |        |
| $SP_4$ | -0.23 | 0.264 | 0.253 | 0.151 | 0.238 | 0.218  | 0.314  | 0.767  | 1      |        |        |        |        |        |        |        |        |
| $SP_5$ | -0.11 | 0.239 | 0.212 | 0.205 | 0.175 | 0.216  | 0.273  | 0.784  | 0.942  | 1      |        |        |        |        |        |        |        |
| $SP_6$ | 0.206 | 0.257 | 0.263 | 0.237 | 0.032 | 0.081  | 0.324  | 0.589  | 0.674  | 0.726  | 1      |        |        |        |        |        |        |
| $SN_1$ | -0.02 | 0.546 | -0.14 | 0.028 | 0.92  | 0.934  | 0.058  | 0.252  | 0.292  | 0.273  | 0.129  | 1      |        |        |        |        |        |
| $SN_2$ | 0.106 | 0.742 | -0.17 | -0.06 | 0.824 | 0.846  | -0.01  | 0.16   | 0.223  | 0.197  | 0.221  | 0.904  | 1      |        |        |        |        |
| $SN_3$ | -0.18 | 0.635 | 0.067 | -0.16 | 0.341 | 0.327  | -0.05  | 0.157  | 0.188  | 0.143  | 0.299  | 0.421  | 0.497  | 1      |        |        |        |
| $SN_4$ | -0.22 | 0.641 | 0.095 | -0.2  | 0.355 | 0.307  | -0.01  | 0.127  | 0.206  | 0.129  | 0.288  | 0.408  | 0.492  | 0.99   | 1      |        |        |
| $SN_5$ | -0.19 | 0.927 | -0.04 | -0.27 | 0.59  | 0.576  | -0.05  | 0.161  | 0.306  | 0.247  | 0.22   | 0.643  | 0.794  | 0.688  | 0.714  | 1      |        |
| $SN_6$ | -0.16 | 0.975 | -0.15 | -0.28 | 0.562 | 0.59   | -0.04  | 0.292  | 0.279  | 0.254  | 0.227  | 0.616  | 0.767  | 0.66   | 0.666  | 0.952  | 1      |

**Fig. 10 Correlation table of Pearson coefficients for all centrality measures**

We observe that almost none of the centrality indices used correlate with any other. This suggests that the rankings introduced by each of the centralities are novel and unique.

From this, we infer that $EC$ is highly correlated with higher values of $SN_i$, which indeed verifies that is a global centrality index, since $SN_i$ focuses on longer closed walks in the network as $i$ increases.

A further observation that deserves attention is the fact that $EC$ does not correlate with $CC$. Also, $DC$ does not correlate with $SC$. This is in stark contrast with earlier work involving the use of centrality measures to address landscape connectivity, where the aforementioned centrality measures were highly correlated with each other [22]. This shows that such correlations are heavily dependent on the structure of the network involved and cannot be generalized to any arbitrary landscape. Hence, a centrality measure, in general, cannot be a substitute for providing rankings intended by another centrality measure. This further urges the use of multiple centrality measures in a network to identify the important vertices.

Using the Newman – Girvan algorithm, the following communities were detected in the network:

Community I: $\{1, 2, 3, 4, 22\}$

Community II: $\{11\}$

Community III: $\{5, 6, 7\}$

Community IV: $\{8, 9, 10, 12, 13, 21, 23\}$

Community V: $\{14, 17, 18, 19, 20\}$

Community VI: $\{15, 16\}$

The fact that 11 appeared as an isolated single-patch community is further supplemented by the observation that it appears in the five least-ranked vertices of every centrality measure used in our analysis, giving further credibility to the rankings of the centrality measures.

## 5. Conclusion

The present work has been developed with objectives to (i) obtain a cost-wise optimal and feasible tiger corridor network, connecting the habitat patches for the tiger in the landscape complex using a replicable computational procedure and (ii) identify the most important habitat patches, along with their underlying community structure so as to focus efforts towards conserving them.

In this paper, we have used Kruskal's algorithm to obtain a minimum spanning tree that could serve as a model framework for a real-world tiger corridor designing in the Central India – Eastern Ghats landscape complex. Then, we subject the graph to centrality analysis in order to identify the potentially important habitat patches. Further, we apply the Newman – Girvan algorithm for detecting communities in the network.

A limitation in the modelling described in the paper is that the corridor designing is based entirely on the structural definition of connectivity, and thus does not take into account some critically vital landscape features such as the biotic factors of availability of prey base and water, in computing the cost matrix. The work is, by choice, kept rudimentary so as to provide a basic computational framework for perceiving a viable structural corridor network design in the focal landscape complex for tigers. A simplifying assumption in the work has been an absence of consideration of multiple possible paths between connected CTHs. We may justify this absence of path redundancy consideration due to two reasons: first, our priority in the paper was to focus on network efficiency over redundancy, and second, the work focuses on estimation of optimal spanning tree connecting the CTHs, rather than inclusion of alternative paths [55]. We are aware that such a simplification is more often not in consonance with the real-world corridor scenario. We however hope that our present effort would make available a computational template for tiger corridor designing, which could certainly be improved upon by incorporating field data from realistic considerations.

The potential corridors identified in the focal landscape complex by the MST are not all least resistance paths, with minimum of the possible costs taken into account for selection of edge by the algorithm, based only on structural considerations, as has been pointed out in the foregoing. These connections, as have been displayed by the output MST, serve merely the purpose of a skeletal design. We would like to emphasize that the result obtained in this work, in order to become useful for wildlife policy considerations, requires fine-tuning through proper validation with actual field data. In this context, it may be worth noting that a recent finding by a team including two of the present authors that was published during writing of this paper, suggest that the estimations of MST as has been obtained in this work is indeed one of the preferred and frequented corridors in actual use by the tiger population in this landscape complex [64].

A secondary type of limitation arises in the use of centrality measures for ranking the habitats; centrality measures do not quantify the difference in importance between different levels of the ranking. So while they can be used to identify potentially important habitats, they provide no information on how important a habitat is, relative to another. It is hoped that fine-grained GIS data, once incorporated appropriately in the designed model, would be able to provide some guidelines to address this limitation.